\documentclass[aps,prl,reprint,groupedaddress]{revtex4-1}

\usepackage{graphicx}
\usepackage{braket}
\usepackage{amsmath}
\usepackage{times}
\usepackage[english]{babel}
\usepackage{amsfonts}
\usepackage{color}
\usepackage[normalem]{ulem}

\newcommand{\nth}{n_{\mathrm{m}}^{\mathrm{th}}}
\newcommand{\kex}{\kappa_{\mathrm{ext}}}
\newcommand{\etain}{\eta_{\mathrm{in}}}

\newcommand{\nf}{n_{\mathrm{m}}}
\newcommand{\etadet}{\eta_{\mathrm{det}}}

\newcommand{\Gammatot}{\Gamma_{\mathrm{tot}}}

\newcommand{\thetaopt}{\theta_0}
\newcommand{\nthd}{n_{\mathrm{l}}}

\newcommand{\Gammaopt}{\Gamma_{\mathrm{opt}}}
\newcommand{\nbath}{n_{\mathrm{m}}^{\mathrm{bath}}}
\newcommand{\nmo}{n_{\mathrm{m}}^0}
\newcommand{\rc}{r_{\mathrm{c}}}
\newcommand{\neffu}{n_{\mathrm{eff}}^{+}}
\newcommand{\neffl}{n_{\mathrm{eff}}^{-}}

\newcommand{\Deltaopt}{\Delta_{0}}
\newcommand{\rin}{r_{\mathrm{in}}}
\newcommand{\nbathimpure}{\tilde{n}_{\mathrm{m}}^{\mathrm{bath}}}

\begin{document}
\title{Sideband Cooling Beyond the Quantum Limit with Squeezed Light}

\author{Jeremy B. Clark, Florent Lecocq, Raymond W. Simmonds, Jos\'{e} Aumentado, John D. Teufel}
\affiliation{National Institute of Standards and Technology, Boulder, CO 80305 USA}

\date{\today}

\begin{abstract}
Quantum fluctuations of the electromagnetic vacuum produce measurable physical effects such as Casimir forces and the Lamb shift \cite{milonni_quantum_1993}.
Similarly, these fluctuations also impose an observable quantum limit to the lowest temperatures that can be reached with conventional laser cooling techniques \cite{marquardt_quantum_2007, wilson-rae_theory_2007}.
As laser cooling experiments continue to bring massive mechanical systems to unprecedented temperatures \cite{teufel_sideband_2011, chan_laser_2011}, this quantum limit takes on increasingly greater practical importance in the laboratory \cite{peterson_laser_2016}.
Fortunately, vacuum fluctuations are not immutable, and can be ``squeezed" through the generation of entangled photon pairs.
Here we propose and experimentally demonstrate that squeezed light can be used to sideband cool the motion of a macroscopic mechanical object below the quantum limit.
To do so, we first cool a microwave cavity optomechanical system with a coherent state of light to within $15\%$ of this limit.
We then cool by more than 2~dB below the quantum limit using a squeezed microwave field generated by a Josephson Parametric Amplifier (JPA).
From heterodyne spectroscopy of the mechanical sidebands, we measure a minimum thermal occupancy of $0.19\pm0.01$ phonons.
With this novel technique, even low frequency mechanical oscillators can in principle be cooled arbitrarily close to the motional ground state, enabling the exploration of quantum physics in larger, more massive systems.
\end{abstract}

\maketitle
Rapid progress in the control and measurement of massive mechanical oscillators has enabled tests of fundamental physics, as well as applications in sensing and quantum information processing \cite{aspelmeyer_cavity_2014}.
The noise performance of these experiments, however, is often limited by thermal motion of the mechanical mode.
Although the most sophisticated refrigeration technologies can be sufficient for cooling high frequency mechanical structures to the ground state \cite{oconnell_quantum_2010, riedinger_non-classical_2016}, observing quantum behavior in lower frequency mechanical systems requires other cooling methods.
Recent efforts using active quantum feedback have been remarkably successful in preparing motional states with low entropies \cite{wilson_measurement-based_2015}.
Thus far, however, only laser cooling techniques similar to those that revolutionized the coherent control of atomic systems \cite{phillips_nobel_1998, wineland_nobel_2013} have yielded thermal occupancies below one quantum \cite{teufel_sideband_2011, chan_laser_2011, peterson_laser_2016}.
Nevertheless, vacuum fluctuations impose a lowest possible temperature that can be achieved using these techniques \cite{marquardt_quantum_2007, wilson-rae_theory_2007}.
This limit is now being encountered in state of the art experiments involving macroscopic oscillators \cite{peterson_laser_2016}.

\begin{figure}
	\includegraphics[width = \columnwidth]{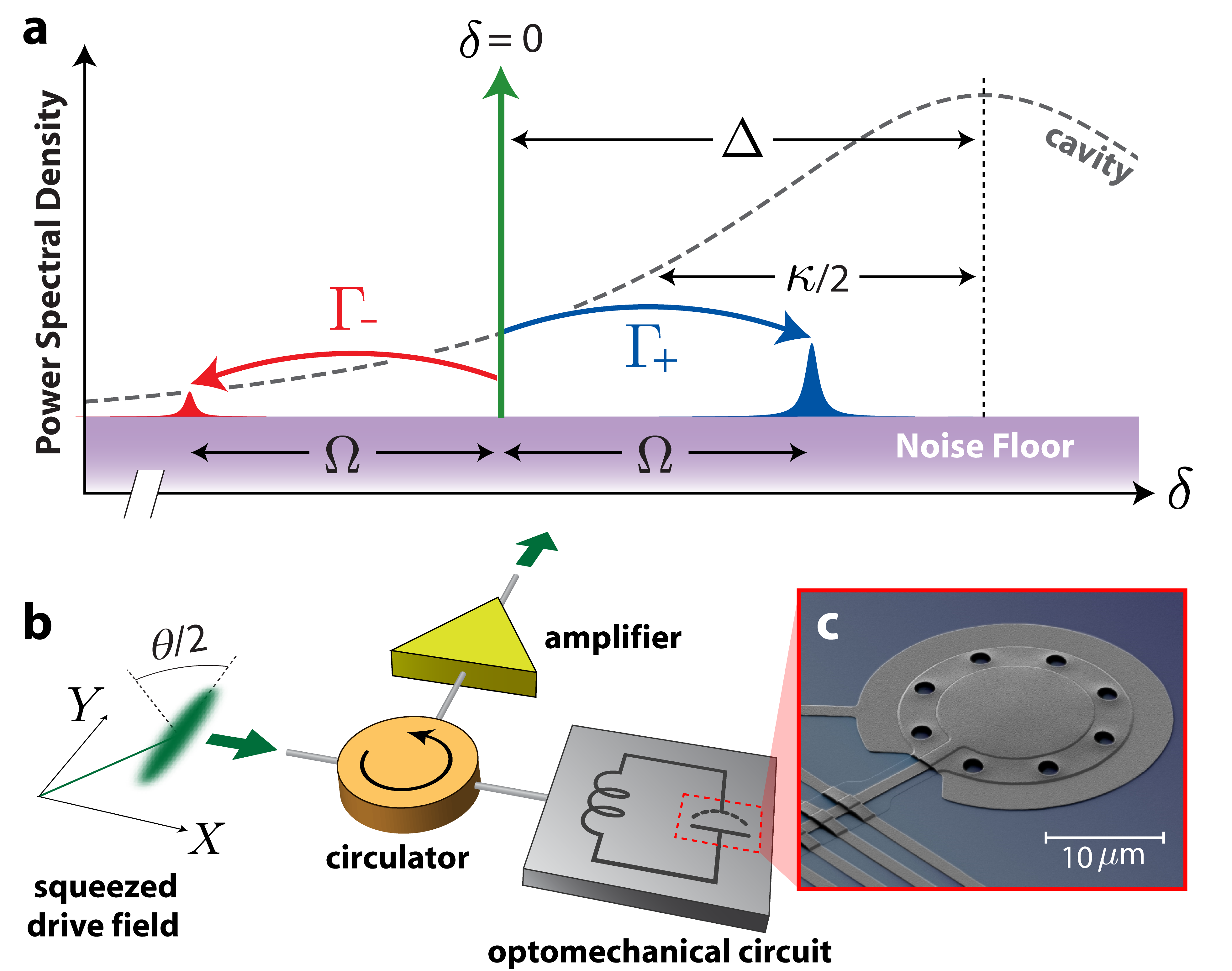}
	\caption{\label{fig:setup}
		Experimental setup.
		\textbf{a}, An optomechanical cavity (full width at half maximum, $\kappa$) is driven at detuning, $\Delta$, with a displaced squeezed state.
		The interaction scatters photons to mechanical sidebands at $\delta=\pm\Omega$, where $\Omega$ denotes the mechanical resonance frequency.
		An anti-Stokes scattering process (upper sideband) carries energy away from the mechanical mode at rate $\Gamma_+$.
		The Stokes scattering rate (lower sideband) heats the mechanical mode at rate $\Gamma_-$.
		\textbf{b}, Microwave implementation.
		A squeezed microwave state of phase $\theta$ drives a resonant circuit consisting of a spiral inductor and vacuum-gap parallel plate capacitor.
		\textbf{c}, False-colored scanning electron micrograph of the mechanically compliant aluminum membrane that forms the top plate of the capacitor.
	}
\end{figure}

The concept of sideband cooling relies on the removal of mechanical energy by scattering incident drive photons to higher frequencies.
In general, however, this photon up-conversion (anti-Stokes) process competes with a down-conversion (Stokes) process that adds energy to the mechanical system.
In cavity optomechanics \cite{aspelmeyer_cavity_2014}, a light-matter interaction arises due to a parametric modulation of an optical cavity's resonance frequency with a mechanical oscillator's position.
When the cavity is driven at detuning $\Delta$ below its resonance frequency, the difference in the cavity's density of states at the mechanical sideband frequencies leads to a dominant anti-Stokes scattering rate (see Fig.~\ref{fig:setup}).
The disparity between the Stokes and anti-Stokes scattering rates grows as the cavity's linewidth, $\kappa$, decreases relative to the mechanical oscillator's resonance frequency, $\Omega$.
Accordingly, optomechanical systems in the ``resolved sideband" limit ($\kappa\ll\Omega$) can be cooled to low temperatures with coherent states of light.
A full quantum analysis, however, shows that vacuum fluctuations always stimulate some degree of Stokes scattering \cite{marquardt_quantum_2007, wilson-rae_theory_2007}, which prevents true ground state cooling.

\begin{figure*}
	\includegraphics[width = \textwidth]{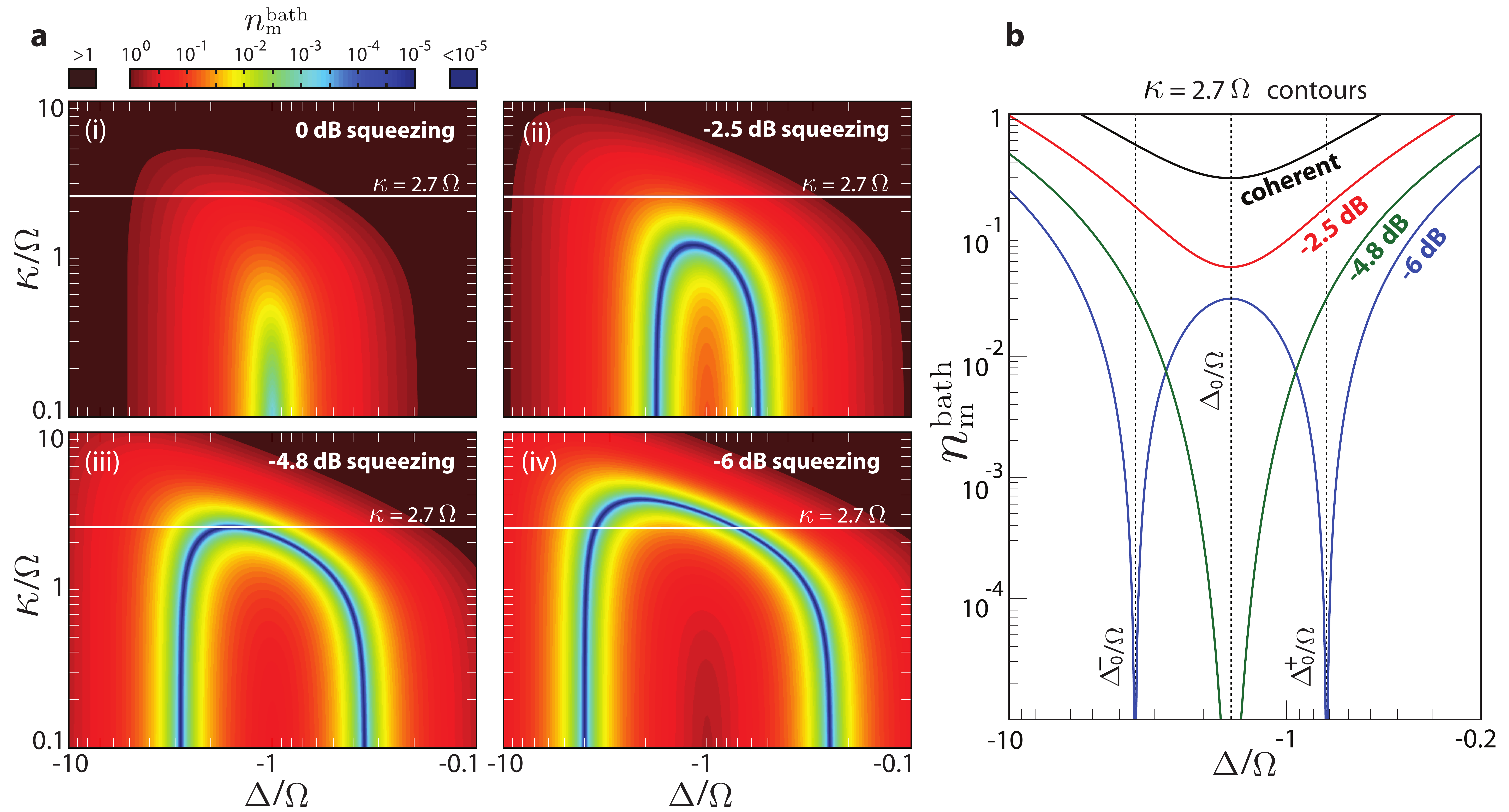}
	\caption{\label{fig:nbathmaps}
		\textbf{a}, Theoretical mechanical bath temperature $\nbath$ (in units of phonons) established by squeezed light for various cavity linewidths, $\kappa$, and drive detunings, $\Delta$, normalized to the mechanical resonance frequency, $\Omega$.
		Panel (i) maps the theory for traditional sideband cooling with a pure coherent state.
		Panels (ii), (iii), and (iv) map the bath temperatures established by -2.5~dB, -4.8~dB, and -6~dB of pure squeezing, respectively.
		The squeezing phase is optimized to $\theta=\thetaopt$ at all points in phase space to consider the lowest possible bath temperatures.
		\textbf{b}, Line cuts of corresponding to our experimental cavity linewidth, $\kappa=2.7~\Omega$.
		These contours are indicated in part (a) by the white horizontal lines.
		The critical squeezing parameter $\rc=0.55$ (Eq.~\ref{eq:rc}) achieves $\nbath=0$ at drive detuning $\Deltaopt$.
		Detunings $\Deltaopt^\pm$ represent the optimal drive detunings for more strongly squeezed states ($r>\rc$).
		At either of these detunings, $\nbath=0$.}
\end{figure*}

Although the sideband cooling limit rests on an intuitive set of assumptions, it is not a fundamental limit.
Proposals to cool below this limit include pulsed cooling schemes \cite{machnes_pulsed_2012, wang_ultraefficient_2011}, dissipative coupling \cite{elste_quantum_2009}, optomechanically-induced transparency \cite{ojanen_ground-state_2014}, and nonlinear interactions \cite{huang_enhancement_2009, lu_squeezed_2015}. 
For atomic laser cooling, it has been proposed \cite{cirac_laser_1993, graham_laser_1991, shevy_laser_1990} that squeezed light can yield an advantage over laser cooling with coherent states.
Here we propose and implement an analogous quantum-enhanced cooling scheme for cavity optomechanical systems.
Using the Heisenberg-Langevin equations, we show that driving an optomechanical cavity with a pure squeezed state of light can coherently null the Stokes scattering process, theoretically eliminating the quantum backaction limit from sideband cooling.
We then implement this squeezing-enhanced cooling method in a microwave cavity optomechanical system \cite{teufel_sideband_2011} to achieve cooling below the quantum limit.

\begin{figure*}
	\includegraphics[width = \textwidth]{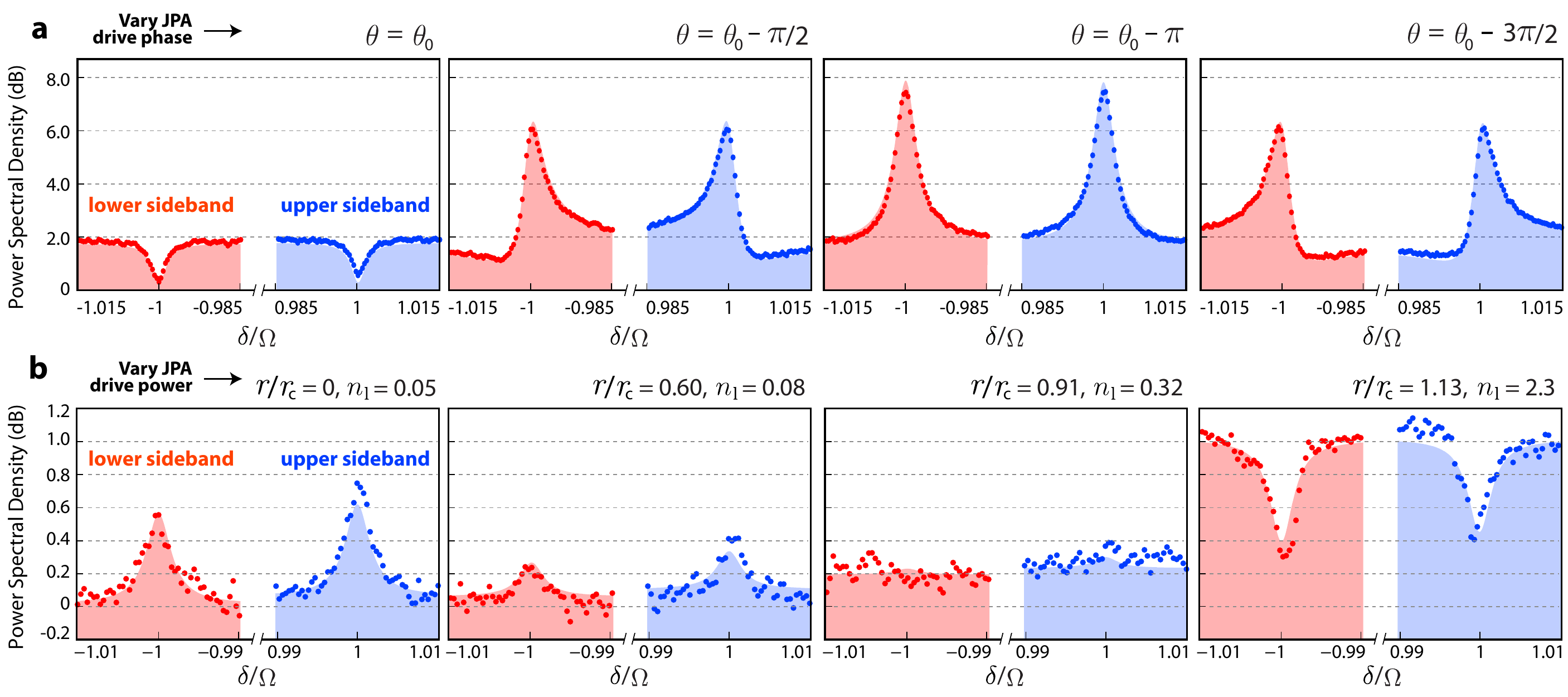}
	\caption{\label{fig:PSD}
		Experimentally measured power spectral densities (normalized to the shot noise limit) of the upper and lower mechanical sidebands for various squeezed states driving the optomechanical cavity at detuning $\Delta=\Deltaopt$.
		The shaded regions show the theoretically predicted behavior.
		\textbf{a}, Measured sidebands as a function of the squeezing phase, $\theta$, for $r/\rc = 1.3$ and $\nthd = 4.2$.
		\textbf{b}, Measured sidebands at $\theta=\thetaopt$ as a function of the applied field's squeezing parameter, $r$.
		The Lorentzian peak becomes a dip when $r$ exceeds $\rc=0.55$, as expected from Eq.~\ref{eq:neff}.
	}
\end{figure*}

The Hamiltonian that governs the interaction between cavity mode $\hat{a}$ and mechanical mode $\hat{b}$ is given by ${\hat{H}_{\mathrm{int}}=-\hbar g(\hat{a}^\dagger+\hat{a})(\hat{b}^\dagger+\hat{b})}$.
Here, $g$ is the parametrically enhanced optomechanical coupling rate, which is proportional to the amplitude of the drive field \cite{aspelmeyer_cavity_2014}.
When driving the system with a pure coherent state, the Stokes and anti-Stokes scattering rates are respectively proportional to $\Gamma_-\times(\nf+1)$ and to $\Gamma_+\times\nf$ in the weak coupling limit ($g\ll\kappa$).
Here, $\nf$ is the equilibrium phonon occupancy of the mechanical mode, and
\begin{equation}
\label{eq:rates}
\Gamma_\pm=\frac{4g^2\kappa}{\kappa^2+4(\Omega\pm\Delta)^2}.
\end{equation}
As $g$ increases, the products ${\Gamma_-\times(\nf+1)}$ and ${\Gamma_+\times\nf}$ tend toward equilibrium, which removes the cooling power of the scattering processes.
Accordingly, the minimum possible phonon occupancy $\nmo$ should satisfy the condition ${\Gamma_+\times\nmo = \Gamma_-\times(\nmo+1)}$, or
\begin{equation}
\label{eq:sblimit}
\nmo = \frac{\Gamma_-}{\Gamma_+-\Gamma_-} = -\frac{\kappa^2+4(\Delta+\Omega)^2}{16\Delta\Omega}.
\end{equation}
The expression for $\nmo$ is minimized at the optimal drive detuning ${\Deltaopt = -\tfrac{1}{2}\sqrt{\kappa^2+4\Omega^2}}$.
Even at this ideal detuning, however, $\nmo$ does not vanish. 

It is convenient to describe the cooling process by modeling the mechanical mode as being coupled to two thermal reservoirs held at different temperatures \cite{marquardt_quantum_2007, wilson-rae_theory_2007}.
One of these reservoirs is provided by the surrounding thermal environment, which results in an initial thermal phonon occupancy, $\nth$.
The coupling rate to this environment is given by the mechanical oscillator's intrinsic decoherence rate, $\Gamma$.
At the same time, the mechanics also couples to an effective thermal reservoir whose temperature is determined by the quantum state of the applied light field.
The coupling rate to this reservoir is given by the optical damping rate, ${\Gammaopt = \Gamma_+ - \Gamma_-}$.
For a coherent state of the drive field, the bath temperature is parameterized by $\nmo$.
The equilibrium phonon occupancy, $\nf$, follows from detailed balance:
\begin{equation}
\label{eq:nf}
\nf = \frac{\Gamma\times\nth + \Gammaopt\times\nmo}{\Gamma + \Gammaopt}.
\end{equation}

The enhancement provided by the squeezing can be quantified by generalizing the expression for $\nmo$ to the case of squeezed drive states.
The resulting bath temperature, $\nbath$, varies with both the strength and phase of the applied squeezing.
We will parameterize the squeezing's strength and phase, respectively, by a squeezing parameter, $r$, and phase, $\theta$ \cite{loudon_squeezed_1987}.
%With this convention \cite{loudon_squeezed_1987}, the light field's amplitude quadrature variance obeys ${\braket{(\Delta\hat{X})^2}=(\cosh2r-\cos\theta\sinh2r)/4}$.
The squeezing phase that minimizes $\nbath$ varies with drive detuning, $\Delta$, according to ${\thetaopt=-4\Delta\kappa/(\kappa^2+4(\Omega^2-\Delta^2))}$.
When driving an optomechanical system at this phase, 
\begin{equation}
\label{eq:nbath}
\nbath=\frac{(\sqrt{\Gamma_-}\cosh r - \sqrt{\Gamma_+}\sinh r)^2}{\Gammaopt}.
\end{equation}
The consequences of Eq.~\ref{eq:nbath} are far-reaching.
While intuition might suggest that the cooling enhancement scales with the squeezing (\textit{i.e.} that $\nbath \propto \nmo \times e^{-2r}$), this is not the case. 
Instead, Eq.~\ref{eq:nbath} reveals that it is always possible to select a finite squeezing parameter, $r$, such that $\nbath$ vanishes.
To do so, $r$ must meet or exceed a critical value, $\rc$, given by
\begin{equation}
\label{eq:rc}
r{_\mathrm{c}}=\frac{1}{2}\sinh^{-1}\bigg(\frac{\kappa}{2\Omega}\bigg).
\end{equation}
This critical squeezing strength is deeply related to a limit in the ``ponderomotive squeezing" produced by off-resonantly driving the cavity with a coherent state (see supplementary information).
 
An optomechanical cavity driven with a pure, critically squeezed state at the traditional optimal detuning, $\Deltaopt$, yields ${\nbath=0}$.
For more strongly squeezed states (${r>\rc}$), the optimal drive detuning bifurcates into a pair of optimal detunings ${\Deltaopt^\pm=-\Omega\cosh(2r)\pm\tfrac{1}{2}\sqrt{4\Omega^2\sinh^2(2r)-\kappa^2}}$.
At either of these detunings, $\nbath=0$.
Figure~\ref{fig:nbathmaps} illustrates this behavior by mapping $\nbath$ as a function of $\kappa/\Omega$ and $\Delta/\Omega$ for several values of the squeezing.
Notably, the plots show how the use of squeezed light pushes colder mechanical bath temperatures into the ``bad cavity" limit ($\kappa\gg\Omega$) where traditional sideband cooling with coherent states is less effective.

Realistic laboratory conditions place practical bounds on the strength and the purity of the squeezing that can be generated to achieve a cooling enhancement.
To address these limitations, we will exploit the freedom to model any impure Gaussian state as a thermal state subject to an ideal entropy-preserving squeezing operation \cite{paris_purity_2003}.
In this picture, the product of the major and minor axes of the light field's noise ellipse obeys ${\braket{(\Delta\hat{X})^2(\Delta\hat{Y})^2}=(1+2\nthd)^2/16}$.
The purity of the squeezing is parameterized by $\nthd$, which denotes the \textit{effective} thermal occupancy of the light field prior to the squeezing operation.
It should be emphasized that any type of impurity introduced to the light field ({e.g.} by loss, parasitic nonlinear processes, \textit{etc.}) will be treated using this model (see supplementary information).

In the presence of impure squeezing, Eq.~\ref{eq:nbath} generalizes to
\begin{align}
\label{eq:nbathtot}
&\nbathimpure=\nbath~+\nonumber\\
&\nthd\times\frac{(\Gamma_++\Gamma_-)\cosh 2r- 2\sqrt{\Gamma_-\Gamma_+}\sinh 2r}{\Gammaopt}.
\end{align}
An examination of Eq.~\ref{eq:nbathtot} reveals that the critical squeezing strength, $\rc$, and the optimal drive detuning(s), $\Deltaopt^{(\pm)}$, remain unaltered when $\nthd>0$.
The impurity of the squeezing does, however, limit the lowest achievable mechanical bath temperature to $\nbathimpure\geq\nthd$.
This inequality saturates to $\nbathimpure=\nthd$ provided that ``strongly" squeezed light fields ($r\geq\rc$) drive the systm at the optimal detunings.
Thus, even impure squeezed states of light can be used to remove the quantum backaction limit from the cooling physics entirely.

To experimentally demonstrate a cooling enhancement, we drive a microwave cavity optomechanical system with a squeezed state generated by a JPA \cite{clark_observation_2016}.
The microwave ``cavity" consists of a vacuum-gap parallel plate capacitor \cite{cicak_low-loss_2010} shunted by a 15~nH spiral inductor, yielding a cavity resonance frequency of $\omega_{\mathrm{c}}=2\pi\times6.4$~GHz.
Our experiments are performed in a dilution cryostat held at $T=37$~mK, resulting in an initial mechanical occupancy of approximately 75 phonons.
For all experiments, the cavity is driven at the traditional optimal detuning, $\Deltaopt$.
Furthermore, we operate in the limit of strong damping where $\Gammaopt=2\pi\times 36$~kHz exceeds the thermal decoherence rate, $\Gamma\times\nth$, by approximately a factor of 30.
As such, the mechanical temperature is primarily determined by the quantum state of the drive field and not by the thermal environment.
The cavity's linewidth ($\kappa=2\pi\times27~$MHz) is comparable to twice the resonance frequency of the aluminum membrane's primary flexural mode ($\Omega =2\pi\times 10.1~\mathrm{MHz}$).
Importantly, however, the cavity's internal loss rate ($\kappa_0<2\pi\times100$~kHz) is small enough relative to $\kappa$ that the effects of internal cavity loss can be neglected (see supplementary information).
% The intrinsic mechanical decay rate is $\Gamma=2\pi\times15$~Hz.

Figure~\ref{fig:PSD}a displays measured power spectral densities of the two mechanical sidebands when the system is driven at various squeezing phases, $\theta$.
The data show that the sidebands generally display a Fano-like spectral lineshape due to the interference of the injected correlations \cite{clark_observation_2016}.
As the squeezing phase passes through $\thetaopt$ (or $\thetaopt-\pi$), however, the sidebands exhibit Lorentzian lineshapes.
These two cases yield extrema in the mechanical mode temperature, which are of primary interest.

Correctly inferring the mechanical temperature from the heterodyne spectra can be subtle.
Although the optimum squeezing phase yields Lorentzian mechanical sidebands, an interference effect persists that distorts the results of standard sideband thermometry.
For the case of uncorrelated drive noise, this effect has been well-characterized experimentally \cite{safavi-naeini_laser_2013, weinstein_observation_2014}.
The effects of correlated noise are more complicated, but they can be handled by introducing effective phonon occupancies for the upper sideband ($\neffu$) and for the lower sideband ($\neffl$).
Relative to the noise floor, the Stokes scattering rate scales as $\Gamma_-\times(\neffl+1)$, whereas the anti-Stokes rate is proportional to $\Gamma_+\times\neffu$.
When $\theta=\thetaopt$, $n_{\mathrm{eff}}^\pm$ are related to the actual phonon occupancy, $\nf$, by 
\begin{equation}
\label{eq:neff}
n_{\mathrm{eff}}^\pm=\nf \mp\nthd\cosh 2r \mp \sinh^2r \pm \sqrt{\frac{\Gamma_\mp}{\Gamma_\pm}}(\tfrac{1}{2}+\nthd)\sinh 2r.\\
\end{equation}
Thus, extracting $\nf$ from the detected sidebands amounts to subtracting these interference terms from the experimentally measured values, $n_{\mathrm{eff}}^\pm$.

\begin{figure}
	\includegraphics[width = \columnwidth]{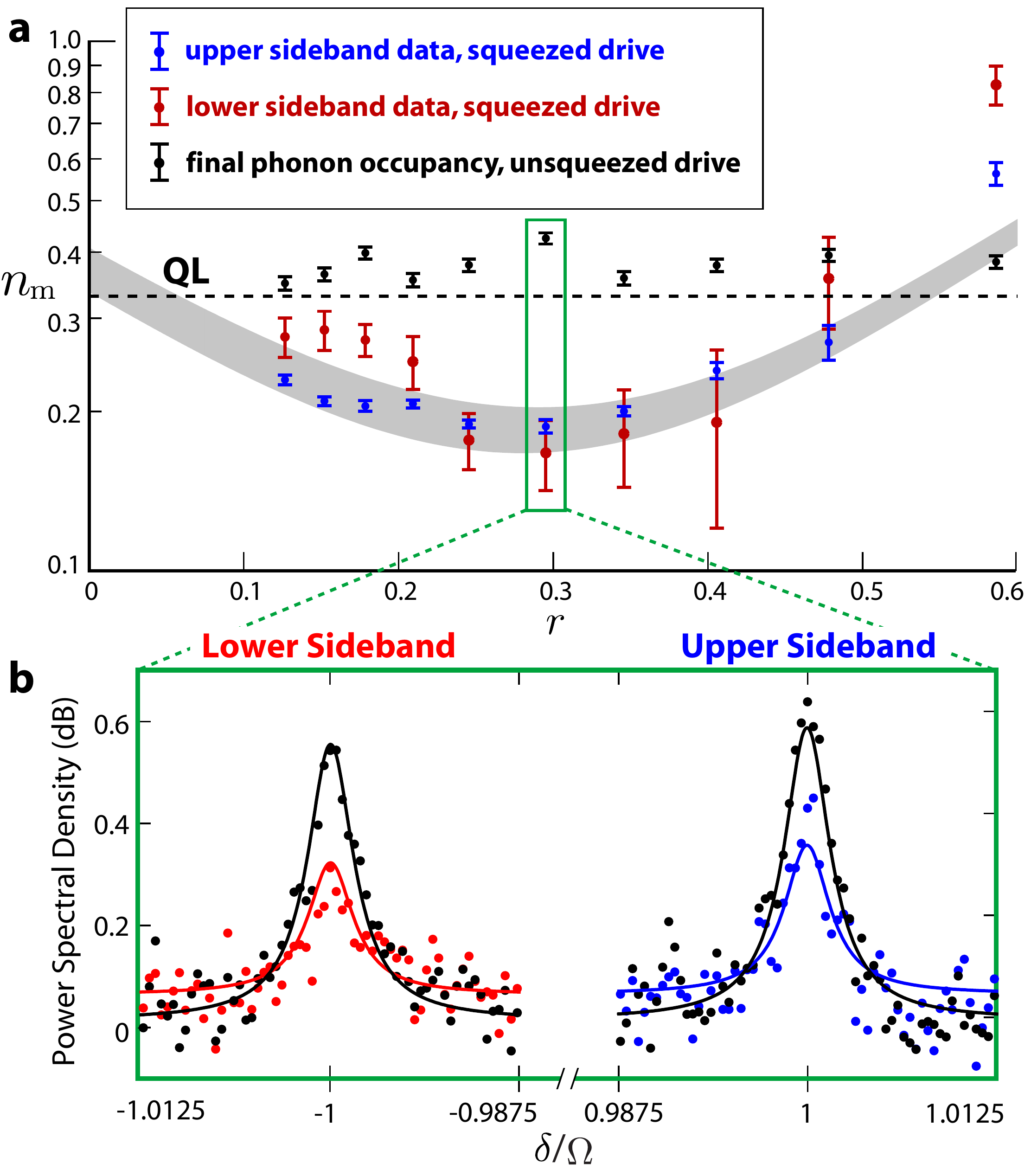}
	\caption{\label{fig:cooling}
		\textbf{a}, Measured phonon occupancy, $\nf$, as a function of the drive's squeezing parameter, $r$.
		The blue and red points correspond to data retrieved from the upper and lower mechanical sidebands, respectively.
		Black points show the measured occupancy with no applied squeezing, indicating sideband cooling to within 15\% of the coherent state ``quantum limit" (dashed line labeled ``QL").
		The gray band illustrates the expected range in the cooling enhancement given independent measurements of the full state of the squeezing.
		\textbf{b}, Power spectral densities of the mechanical sidebands (which are normalized to the shot noise limit) corresponding to the data taken at the coldest mechanical temperature, $\nf=0.19$.
		The indicated uncertainties specify the standard deviation of the mean obtained over 8 experiments.
	}
\end{figure}

Equation~\ref{eq:neff} contains an important signature of the onset of critical squeezing in the strong damping limit (${\Gammaopt\gg\Gamma\times\nth}$).
For any $\nthd\geq0$, the effective phonon occupancies reduce to $n_{\mathrm{eff}}^+=0$ and $n_{\mathrm{eff}}^-=-1$ when $r=\rc$ and $\Delta=\Deltaopt$.
In other words, a critically squeezed state nulls both mechanical sidebands to the noise floor, regardless of the squeezed state's purity.
Figure~\ref{fig:PSD}b illustrates this effect as $r$ is increased through $\rc=0.55$.
When $r>\rc$ (and the detuning is held to $\Delta=\Deltaopt$), the mechanical sidebands dip below the heterodyne noise floor.
In a manner analogous to active feedback cooling \cite{wilson_measurement-based_2015}, these dips signal the onset of mechanical heating.

In Fig.~\ref{fig:cooling}, we use heterodyne spectroscopy of both mechanical sidebands to measure the thermal occupancy of the mechanical mode \cite{purdy_optomechanical_2015, underwood_measurement_2015, peterson_laser_2016}.
Starting first with a coherent state, we cool the mechanical mode to within 15\% of the quantum limit, ${\nmo=0.33}$.
Without changing $\Gammaopt$, we then increase the squeezing parameter, $r$, by increasing the pump power to the JPA (while holding $\theta=\thetaopt$).
As shown in Fig.~\ref{fig:cooling}b, the heights of the measured sidebands remain nearly identical for any value of $r$.
Using Eq.~\ref{eq:neff}, we extract $\nf$ from the sidebands as a function of the squeezing strength.
These results are then compared against theory, which is represented by the gray zone plotted in Fig.~\ref{fig:cooling}a.
As expected, our results show that the cooling enhancement improves with the strength of the correlations for smaller values of the squeezing parameter ($r<0.25$).
As $r$ is increased, however, the limited microwave transmittance ($\etain=57 \pm 2\%$) between the JPA and the optomechanical cavity leads to a degradation in the purity of the squeezing driving the system.
This leads to an inverse relationship between $r$ and $\nthd$ and a minimum mechanical occupancy $\nf=0.19\pm0.01$ at $r=0.3$.

Our data demonstrate that squeezed light can be used to cool a cavity optomechanical system below the quantum limit.
Analytical expressions for the mechanical mode temperature indicate that the primary cooling bottleneck is set by the purity of the applied squeezing.
A natural extension of this effort would be to parametrically swap an itinerant squeezed light field into the mechanical state \cite{palomaki_coherent_2013, aspelmeyer_cavity_2014}, promising mechanically squeezed states of unprecedented strength.
Looking forward, this generic technique could immediately improve the cooling of any generic cavity optomechanical system, even allowing true ground state cooling of systems outside of the resolved sideband regime.

\begin{acknowledgments}
This work was supported by NIST and the DARPA QuASAR program.
J.B. Clark acknowledges the NIST National Research Council Postdoctoral Research Associateship Program for its financial support.
\end{acknowledgments}
\bibliography{sqz_cool_bib_scrub}
\clearpage

% % % % % % % % % % % % % % % % % % % % % % % % % % % % % % % % % % % % % % % % % % % % % % % % % % % % % % % % % % % % % % % % % % % % % % % % % % % % % % % % % % % % % % % % % % % % % %

\onecolumngrid
\begin{center}
	{\large\textbf{Supplementary Information}}
\end{center}
\twocolumngrid
\section{Supplementary Methods}
\subsection{Microwave Set-up}
Figure~\ref{fig:setup_complete} depicts the thermalization stages of the dilution cryostat used to perform the experiments.
The room temperature microwave set-up (not pictured) was identical to that detailed in \cite{clark_observation_2016}, with only the heterodyne detection pathways selected.
We note an increase in the cold stage attenuation of the microwave drive lines to the Josephson Parametric Amplifier (JPA) and to the optomechanical (OM) circuit.
This change was intended to minimize any observable thermal noise on the microwave field driving the optomechanical circuit, which we estimate to be less than 0.05 thermal photons.
The initial thermal occupancy of the mechanical mode (approximately 75 phonons) agrees with the measured base temperature ($T=37~\mathrm{mK}$) of the dilution cryostat.
The noise performance of the microwave heterodyne detection is set by a cryogenic, HEMT amplifier.
We measure a detection chain noise temperature of $T_{\mathrm{N}}=4.6~\mathrm{K}$, which corresponds to an effective heterodyne detection efficiency of $\etadet=6.5\%$.

\subsection{Measurement of $\nf$}
We describe the procedure used to retrieve the equilibrium phonon occupancy, $\nf$, from the measured mechanical sidebands.
When the cavity is driven with squeezed light, the power spectral densities of the sidebands generally display ``Fano" lineshapes (see Fig.~\ref{fig:PSD}a).
The exact shapes of the sidebands depend on the phase of the applied squeezing, $\theta$ (defined in Fig.~\ref{fig:setup}).
At the two special squeezing phases $\theta = \thetaopt$ and $\theta=\thetaopt-\pi$, however, these sidebands exhibit Lorentzian lineshapes.
Again, $\thetaopt$ denotes the squeezing phase that yields optimal sideband cooling:
\begin{equation}
\label{eq:thetaopt}
\thetaopt=\tan^{-1}\bigg(\frac{-4\Delta\kappa}{\kappa^2+4(\Omega^2-\Delta^2)}\bigg).
\end{equation}
In contrast, the squeezing phase $\theta=\thetaopt-\pi$ leads to maximal heating of the mechanical mode.

\begin{figure}
	\includegraphics[width = \columnwidth]{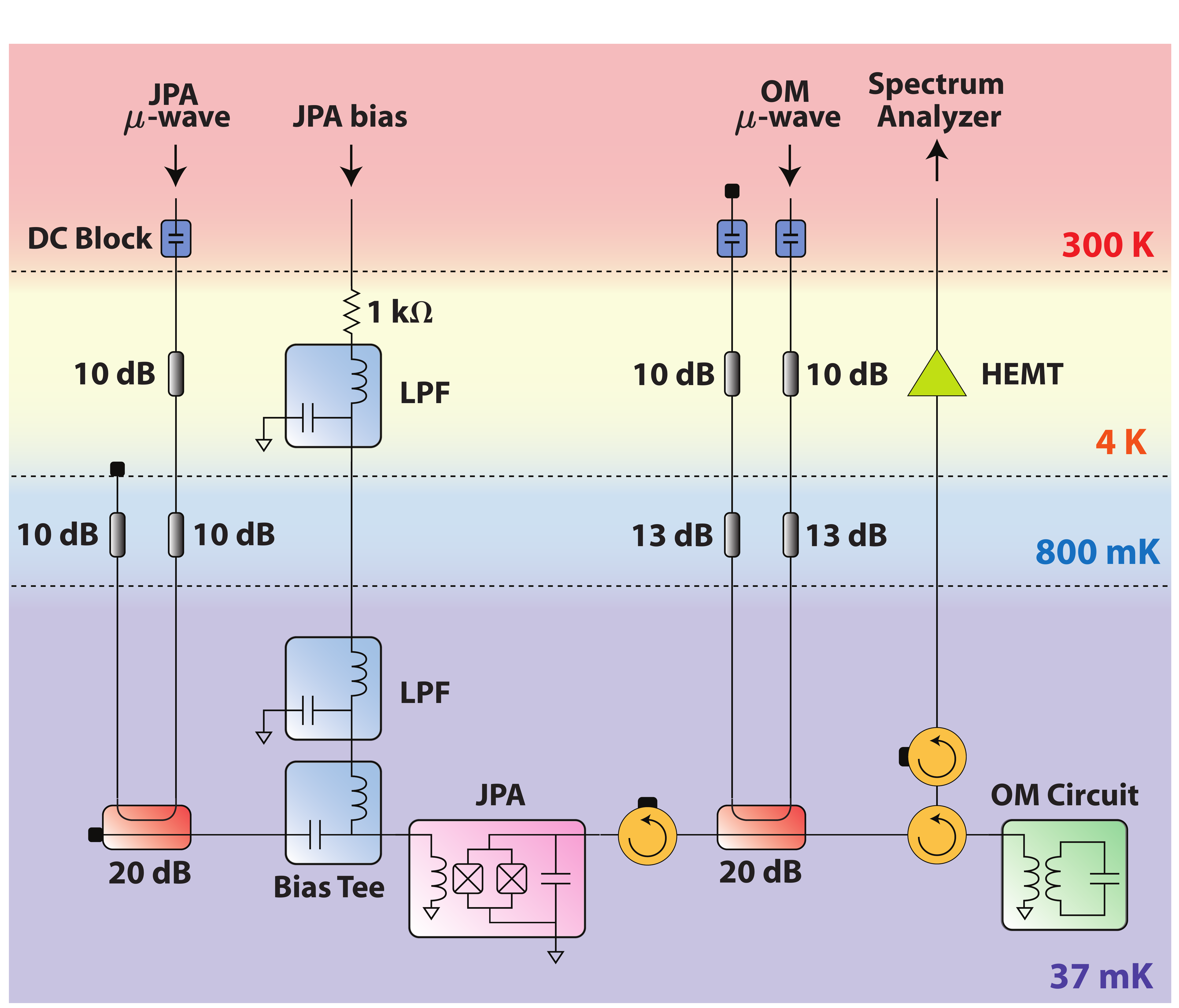}
	\caption{\label{fig:setup_complete}
		Cryogenic measurement set-up.
		LPF = low-pass filter, JPA = Josephson Parametric Amplifier, OM = optomechanics, HEMT = high-electron-mobility transistor amplifier.
		The components shaded in gray signify broadband microwave attenuators.
		The rectangular components shaded red indicate directional couplers.
	}
\end{figure}

The fact that the mechanical sidebands are Lorentzians at these two phases allows a ``naive" application of typical sideband thermometry techniques.
At either of these squeezing phases, the heights of the Lorentzian mechanical sidebands are proportional to
\begin{align}
\label{eq:neff1}
\mathrm{Upper~sideband:}~~&\Gamma_+\times\neffu\\
\label{eq:neff2}
\mathrm{Lower~sideband:}~~&\Gamma_-\times(\neffl+1),
\end{align}
where the heights are taken with respect to the heterodyne noise floor (which rises with the squeezing).
Here, $n_{\mathrm{eff}}^\pm$ represent the effective thermal occupancies of the mechanical mode retrieved from the upper (+) and lower (-) sidebands.
It should be emphasized that these expressions assume a strongly overcoupled cavity.
Additionally, the widths of the Lorentzian sidebands (${\Gammatot=\Gamma_+-\Gamma_-+\Gamma}$) do not change with the squeezing .
$\Gamma_\pm$ are defined in Eq.~\ref{eq:rates} of the main text.

In general, $\neffl\neq\neffu$, and ${n_{\mathrm{eff}}^\pm\neq\nf}$.
Therefore, standard sideband thermometry does not yield the correct mechanical temperature when the system is driven with squeezed light.
Fortunately, the corrections to $n_{\mathrm{eff}}^\pm$ are straightforward to apply.
As explained in the main text, $n_{\mathrm{eff}}^\pm$ are related to the actual mechanical occupancy, $\nf$, by
\begin{align}
\label{eq:neffu_mod}
n_{\mathrm{eff}}^+&=\nf-\nthd\cosh 2r-\sinh^2r\pm\sqrt{\frac{\Gamma_-}{\Gamma_+}}(\tfrac{1}{2}+\nthd)\sinh 2r\\
\label{eq:neffl_mod}
n_{\mathrm{eff}}^-&=\nf+\nthd\cosh 2r+\sinh^2r\mp\sqrt{\frac{\Gamma_+}{\Gamma_-}}(\tfrac{1}{2}+\nthd)\sinh 2r,
\end{align}
where the top (bottom) sign corresponds to the case where ${\theta = \thetaopt}$ (${\theta = \thetaopt - \pi}$).
In Eqs.~\ref{eq:neffu_mod} and \ref{eq:neffl_mod}, $r$ and $\nthd$ respectively denote the squeezing parameter for an \textit{ideal}, unitary squeezing process applied to an initial thermal state of occupancy $\nthd=\mathrm{Tr}(\hat{\rho}_a\hat{a}^\dagger\hat{a})$.
As emphasized in the text, $r$ and $\nthd$ are used to describe any impure Gaussian state, regardless of the origins of the impurity.

Before using Eqs.~\ref{eq:neffu_mod} and \ref{eq:neffl_mod} to infer $\nf$, we retrieve system parameters $g$, $\kappa$, and $\Omega$ by interrogating the optomechanical system with an unsqueezed microwave field using standard techniques.
With these parameters determined, we vary the state of the squeezing by changing the amplitude and phase of the pump driving the JPA (see Fig.~\ref{fig:setup_complete}).
Equations~\ref{eq:neffu_mod} and \ref{eq:neffl_mod} make it clear that $n_{\mathrm{s}}^\pm$ vary as a function of $\kappa$, $\Omega$, $\Delta$, and $\theta$, which are known.
They also vary as a function of $r$ and $\nthd$, which we treat as unknown quantities.

The unknown parameters $r$ and $\nthd$ are uniquely determined by two measurements of $n_{\mathrm{eff}}^+$ (or of $n_{\mathrm{eff}}^-$) at the two squeezing phases of interest, $\theta = \thetaopt$ and $\thetaopt-\pi$.
Since $r$ and $\nthd$ also specify $\nf$ (in conjunction with the known parameters $g$, $\kappa$, $\Omega$, $\Delta$, and $\theta$), we are able to retrieve the mechanical temperature without presupposing the state of the squeezing.
In other words, the measurements of each sideband reveal both the state of the squeezing driving the mechanical mode and the mechanical mode's phonon occupancy.
These measurements were repeated for both the upper and lower sidebands at each value of $r$ in Fig.~\ref{fig:cooling}, yielding two independently retrieved values of $\nf$ (one per sideband).
These results were compared against separate measurements of the squeezing (from homodyne tomography and from QND measurements \cite{clark_observation_2016}), which were used to compute the gray theory zone in Fig.~\ref{fig:cooling}.

\section{Supplementary Discussion}
This section will augment important theoretical ideas that were not discussed at length in the main text.
In particular, we will develop an intuition for the existence of a critical squeezing strength required for ground state cooling.
Additionally, we will discuss several effects that limit the cooling enhancement provided by the squeezing.
These results can be straightforwardly derived using the steady-state solutions of the Heisenberg-Langevin equations.
We therefore begin by specifying the assumptions used to solve those equations.

The Hamiltonian that describes the coupling of the cavity mode $\hat{a}$ to the mechanical mode $\hat{b}$ is given by \cite{aspelmeyer_cavity_2014}
\begin{equation}
\hat{H}_{\mathrm{int}}=-\hbar g_0\hat{a}^\dagger\hat{a}(\hat{b}^\dagger+\hat{b}),
\end{equation}
where $g_0$ represents the vacuum optomechanical coupling rate.
As expressed in the main text, however, we take the common approach of working in a linearized regime in which a coherent buildup of cavity photons ${\braket{\hat{a}^\dagger\hat{a}}=|\alpha|^2}$ (\textit{i.e.} the ``stick" in the ``ball and stick" picture of a Gaussian state) enhances the coupling rate according to $g=\alpha\times g_0$.
The cavity and mechanical field quadrature fluctuations then couple together according to
\begin{equation}
\hat{H}^{\mathrm{lin}}_{\mathrm{int}}=-\hbar g(\hat{a}^\dagger+\hat{a})(\hat{b}^\dagger+\hat{b}).
\end{equation}

Under the linearized coupling assumption, the evolution of the fields is described by coupled Heisenberg-Langevin equations
\begin{align}
\label{eq:HL1}
\dot{\hat{a}}&=-(\tfrac{\kappa}{2}-i\Delta)\hat{a} +ig(\hat{b}^\dagger+\hat{b})+\sqrt{\kex}\hat{\xi}^{\mathrm{ext}}_a+\sqrt{\kappa_0}\hat{\xi}^{\mathrm{int}}_a\\
\label{eq:HL2}
\dot{\hat{b}}&=-(\tfrac{\Gamma}{2}+i\Omega)\hat{b} +ig(\hat{a}^\dagger+\hat{a})+\sqrt{\Gamma}\hat{\xi}_b,
\end{align}
where $\{\hat{\xi}^i_a\}$ and $\hat{\xi}_b$ represent environmental noise operators for modes $\hat{a}$ and $\hat{b}$.
These noise operators couple to the cavity and mechanical fields at rates $\{\kappa_i\}$ and $\Gamma$, respectively.
$\hat{\xi}_a^{\mathrm{ext}}$ denotes the noise environment set by the squeezing.
$\kex$ parameterizes the coupling rate to the feed line carrying the squeezed field.
The cavity's internal dissipation is modeled by introducing an ``internal" noise operator, $\hat{\xi}_a^{\mathrm{int}}$, which is assumed to be at zero temperature.
The mechanical bath operators, $\hat{\xi}_b$, are assumed to satisfy the characteristics of a thermal state, with an average thermal occupancy ${\nth=\mathrm{Tr}(\hat{\rho}_b\hat{\xi}_b^\dagger\hat{\xi}_b)}$.
All baths are assumed to be Markovian (\textit{i.e.} memoryless).
Finally, the input-output relation
\begin{equation}
\hat{\xi}^{\mathrm{ext,~out}}_a=-\hat{\xi}^{\mathrm{ext}}_a+\sqrt{\kex}\hat{a}
\end{equation}
is used to evaluate the heterodyne spectrum of the reflected drive field, $\hat{\xi}^{\mathrm{ext,~out}}_a$.

It should be noted that the solutions to Eqs.~\ref{eq:HL1} and \ref{eq:HL2} contain a well-known ``optical spring" effect:
\begin{align}
&\Omega_{\mathrm{tot}}=\Omega\nonumber\\
&+g^2\bigg(\frac{\Delta-\Omega}{\kappa^2/4+(\Delta-\Omega)^2}+\frac{\Delta+\Omega}{\kappa^2/4+(\Delta+\Omega)^2}\bigg).
\end{align}
In the main text, all references to the mechanical resonance frequency, $\Omega$, actually refer to $\Omega_{\mathrm{tot}}$.
For our experimental parameters, however, this distinction is not crucial since the optical spring amounts to about 1\% correction at our highest applied drive powers.

\subsection{Intuition for the Critical Squeezing}
One of the most counterintuitive aspects of the cooling enhancement is the requirement of a minimum level of applied squeezing to achieve ground state cooling.
In this section, we show that this critical level of squeezing is intimately related to the minimum level of ``ponderomotive" squeezing \cite{fabre_quantum-noise_1994, mancini_quantum_1994, brooks_non-classical_2012, purdy_strong_2013, safavi-naeini_squeezed_2013} that can be generated by driving the optomechanical cavity with a coherent state in the strong damping limit (${\Gammaopt\gg\Gamma\times\nth}$).
To sketch this picture, we will derive the limit of the ponderomotive squeezing that can be achieved for off-resonant coherent state drives.
The critical squeezing parameter, $\rc$, and bifurcated detuning solutions, $\Deltaopt^\pm$, can be derived from this limit.
The discussion will conclude with a calculation of the remaining ponderomotive squeezing after the optomechanical system is seeded with a squeezed state of arbitrary strength, $r$.

\subsubsection{Ponderomotive Squeezing}
\begin{figure}
	\includegraphics[width = \columnwidth]{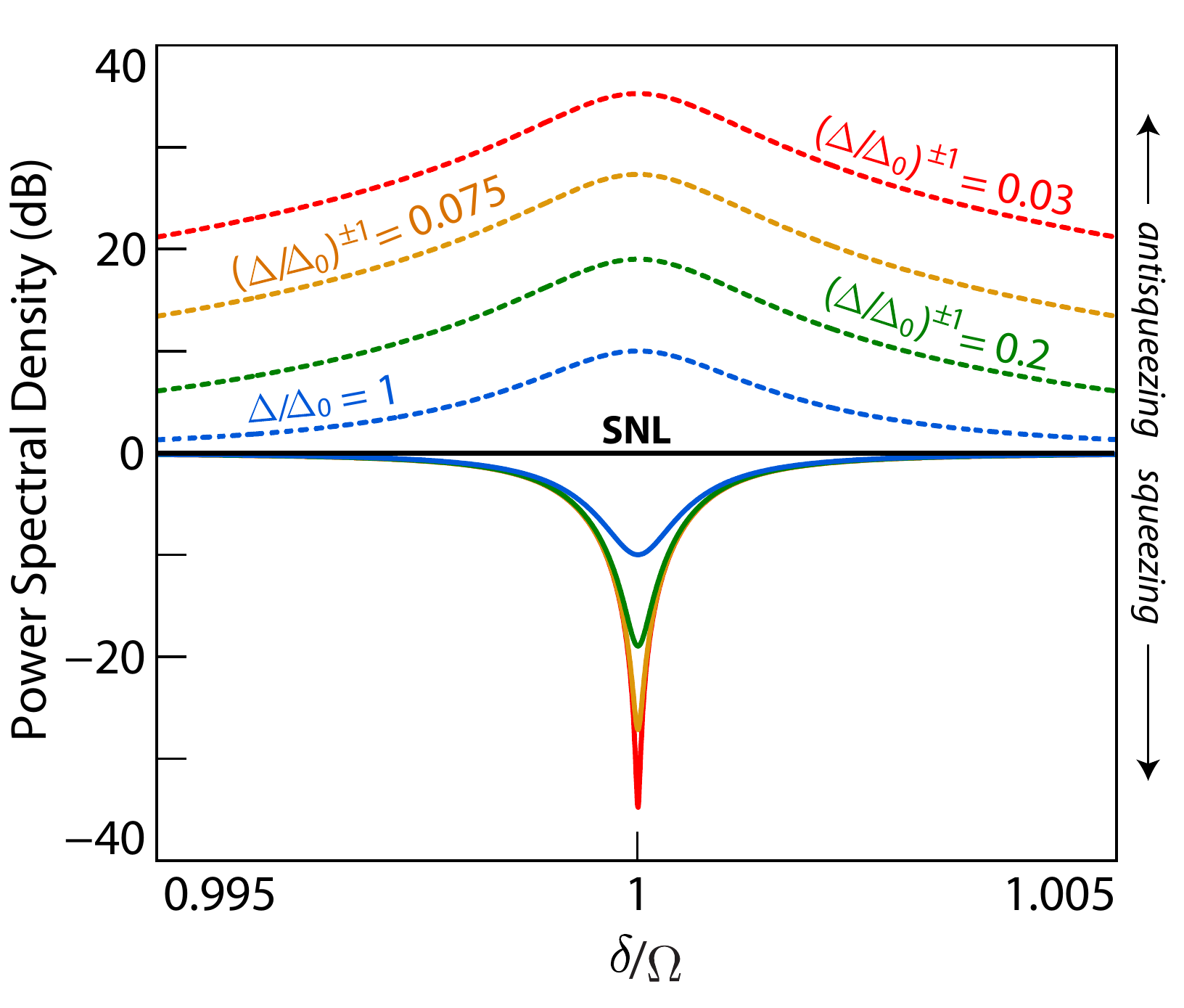}
	\caption{\label{fig:pond_sqz}
		Calculated power spectral density of the ponderomotive squeezing (solid lines) and antisqueezing (dashed lines) that would be detected by an ideal homodyne receiver.
		SNL denotes the shot noise limit (vacuum fluctuations).
		The indicated mechanical resonance frequency, $\Omega$, includes optical springing effects.
		The color of each curve denotes the resulting power spectral density produced by either of two coherent state pump detunings satisfying ${(\Delta/\Deltaopt)^{\pm1}=\mathrm{constant}}$.
		Here, $\Deltaopt$ denotes the traditional optimal drive detuning when sideband cooling with coherent states (Eq.~\ref{eq:Deltaopt}).
		All curves assume a mechanical quality factor $Q_{\mathrm{m}}=5\times10^5$, a constant (and strong) optical damping rate ${\Gammaopt=\Gamma_+-\Gamma_-=10^3\times\Gamma}$, a lossless optomechanical cavity, and zero temperature.
		Additionally, a normalized cavity linewidth $\kappa/\Omega \approx2.85$ has been assumed, which yields -10~dB of ponderomotive squeezing when $\Delta=\Deltaopt$ (see Eq.~\ref{eq:OM_sqz2}).
	}
\end{figure}

\begin{figure*}
	\includegraphics[width = \textwidth]{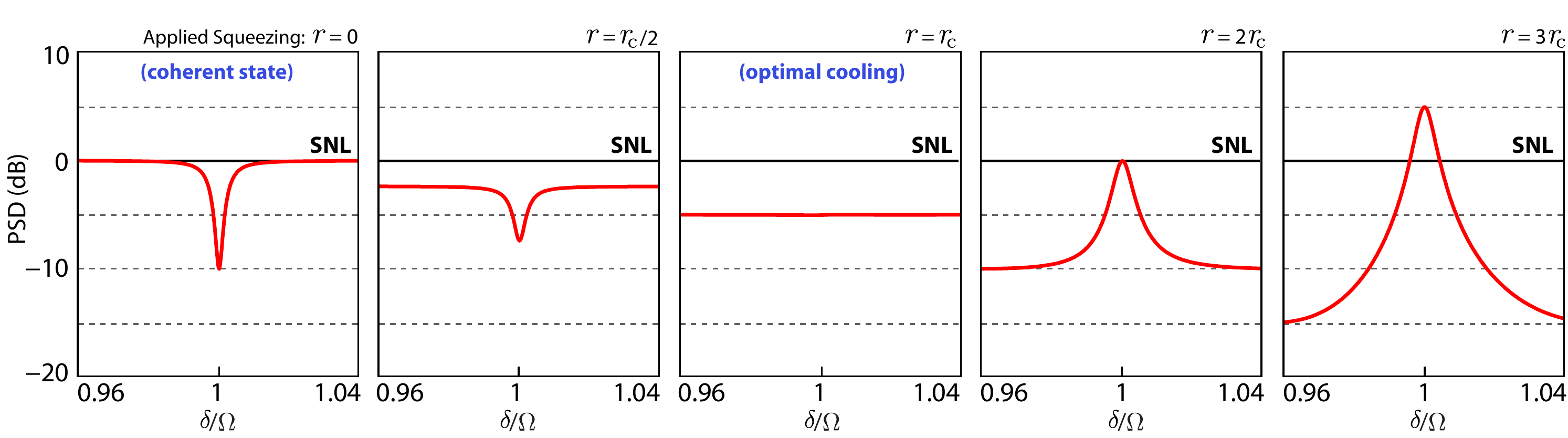}
	\caption{\label{fig:rcrit}
		Calculated power spectral density (PSD) of the reflected drive field as would be detected by an ideal homodyne receiver.
		The field quadrature being plotted is that which is maximally squeezed (ponderomotively).
		SNL denotes the shot noise limit (vacuum fluctuations).
		All panels assume the same conditions as Fig.~\ref{fig:pond_sqz}, with the exceptions of a constant drive detuning ($\Delta=\Delta_0$) and a constant drive cooperativity ${C=\tfrac{4g^2}{\kappa\Gamma}=5000}$.
		Each panel assumes a different level of injected squeezing, though the optimal squeezing phase $\theta=\thetaopt$ (Eq.~\ref{eq:thetaopt}) is assumed throughout.
		The strength of the injected squeezing is parameterized by $r$, which has been expressed in units of the critical squeezing parameter, $\rc$.
		The critical squeezing for this cavity is $-5$~dB.
	}
\end{figure*}

The interaction between an applied coherent state and an optomechanical cavity squeezes the reflected light field near the mechanical resonance frequency, $\Omega$ (which is taken to include ``optical springing" effects).
Typically, such ponderomotive squeezing is generated using a resonant coherent drive tone ($\Delta=0$) since, in this case, the resulting squeezing scales with the applied drive power.
Nevertheless, ponderomotive squeezing persists even in the case where the drive field is tuned below cavity resonance (\textit{i.e.} when $\Delta<0$).
In this case, however, the squeezing does not scale arbitrarily high with pump power.
Instead, at any red drive detuning $\Delta<0$, the state's squeezing parameter $r_{\mathrm{OM}}$ saturates to
\begin{equation}
\label{eq:OM_sqz}
\sinh (r_{\mathrm{OM}})=2\sqrt{\dfrac{\Gamma_-\Gamma_+}{\Gammaopt^2}},
\end{equation}
where $\Gamma_\pm$ are defined in Eq.~\ref{eq:rates}.
Equation~\ref{eq:OM_sqz} assumes negligible thermal mechanical noise, such that ${\Gammaopt\gg\nth\times\Gamma}$.
At zero temperature, Eq.~\ref{eq:OM_sqz} also holds under the weaker condition that $\Gammaopt\gg\Gamma$.

The expression for $r_{\mathrm{OM}}$ in Eq.~\ref{eq:OM_sqz} confirms that the squeezing parameter does not depend on the applied pump strength (which is parameterized by the linearized coupling rate, $g$).
Indeed, Eq.~\ref{eq:OM_sqz} can be more suggestively expressed in terms of the coupling-independent resolved sideband cooling limit, $\nmo$:
\begin{equation}
\label{eq:OM_sqz2}\sinh(r_{\mathrm{OM}})=2\sqrt{\nmo\times(\nmo+1)}.
\end{equation}
Equation~\ref{eq:OM_sqz2}, shows that the ponderomotive squeezing increases under conditions where resolved sideband cooling is \textit{less} effective.
This observation will explain the need for stronger levels of injected squeezing for ground state cooling of systems in the ``bad cavity limit" ($\kappa/\Omega\gg1$).
It will also explain why more strongly squeezed states are needed for drive detunings away from the optimal coherent state drive detuning
\begin{equation}
\label{eq:Deltaopt}
\Deltaopt=-\frac{1}{2}\sqrt{\kappa^2+4\Omega^2}.
\end{equation}

To gain further intuition for Eqs.~\ref{eq:OM_sqz} and \ref{eq:OM_sqz2}, it will be useful to define the normalized drive detuning
\begin{equation}
\tilde{\Delta}\equiv\Delta/\Deltaopt.
\end{equation}
In terms of $\tilde{\Delta}$, $r_{\mathrm{OM}}$ takes the following form:
\begin{equation}
\label{eq:OM_sqz3}
\sinh (r_{\mathrm{OM}})=\frac{1}{2}
\sqrt{\bigg(\tilde{\Delta}+\frac{1}{\tilde{\Delta}}\bigg)^2\bigg(\dfrac{\kappa}{2\Omega}\bigg)^2+\bigg(\tilde{\Delta}-\dfrac{1}{\tilde{\Delta}}\bigg)^2}.
\end{equation}
Equation~\ref{eq:OM_sqz3} reveals several results:
\begin{align}
\label{eq:point2} r_{\mathrm{OM}}(\tilde{\Delta}=1)&=2\rc.\\
\label{eq:point1} r_{\mathrm{OM}}(\tilde{\Delta})&\geq2\rc.\\
\label{eq:point3} r_{\mathrm{OM}}(\tilde{\Delta})&=r_{\mathrm{OM}}(\tilde{\Delta}^{-1}).
\end{align}
Equations~\ref{eq:point2} and \ref{eq:point1} show that the critical squeezing parameter, $\rc$, emerges directly from considering the ponderomotive squeezing without any consideration of ground state cooling.
The symmetry of the squeezing with respect to $\tilde{\Delta}$ (Eq.~\ref{eq:point3}) accounts for the behavior displayed in Fig.~\ref{fig:pond_sqz}.
The next subsection will help elucidate why this symmetry also accounts for the symmetric bifurcation behavior predicted in Fig.~\ref{fig:nbathmaps}a of the main text.

\subsubsection{Optomechanical Response to the Injected Squeezing}
As discussed in the main text, suppressing the lower (Stokes) mechanical sideband opens the possibility of true ground state cooling.
To show when this condition is met, we calculate the expected power spectral density of an ideal homodyne detection of the squeezed drive field after its interaction with a cavity optomechanical system.
The calculation confirms that the power spectral density at the mechanical resonance frequency flattens when the system is driven with a critically squeezed state.

When driving an optomechanical system with a squeezed state at the optimum squeezing phase $\theta=\thetaopt$ (Eq.~\ref{eq:thetaopt}), the quadrature amplification of the incident light field driving the optomechanical cavity is completely out of phase with the amplification provided by the JPA.
Figure~\ref{fig:rcrit} illustrates this behavior.
For example, a weakly squeezed (${r<\rc}$) light field driving the optomechanical system is expected to yield a net reduction in the level of ponderomotive squeezing near $\delta=\Omega$.
By increasing the level of injected squeezing (\textit{i.e.} squeezing from the JPA) to the point that $r=2\rc$, the ponderomotive squeezing at the mechanical sideband is reamplified by the optomechanical system back to the shot noise limit (SNL).
Stronger injected squeezed fields ($r>2\rc$) become further amplified above the SNL near $\delta=\Omega$.

In addition to reversing the deamplification of the optomechanical system, the injection of squeezed light also suppresses the noise floor away from the mechanical sideband frequency.
When ${r=\tfrac{r_{\mathrm{OM}}}{2}=\rc}$, this suppression of the noise floor exactly meets rising noise level at $\delta=\Omega$, which flattens the reflected drive field's power spectral density.
At this critical point, the squeezed field scatters from the cavity as if it were free of any optomechanical interaction (including a frequency-dependent rotation of the squeezing phase).

When an optomechanical cavity produces strong ponderomotive squeezing, it is necessary to inject a correspondingly strong level of externally generated squeezing (in our case, from the JPA) to suppress the mechanical sidebands.
In light of Eq.~\ref{eq:OM_sqz2}, it follows that sideband cooling optomechanical devices in the bad cavity limit (where $\nmo$ is large) also requires stronger levels of injected squeezing.
This observation accounts for the behavior exhibited in Fig.~\ref{fig:nbathmaps}, which shows that squeezing pushes colder bath temperatures into the bad cavity limit.
Similarly, drive detunings $\Delta\neq\Deltaopt$ that produce stronger levels of ponderomotive squeezing require more strongly squeezed fields ($r>\rc$) to cool the mechanical mode to the ground state.
This explains the bifurcation behavior of the optimal drive detunings (see Fig.~\ref{fig:nbathmaps}).
Indeed, solving Eq.~\ref{eq:OM_sqz3} for the detuning roots reproduces the detuning solutions ${\Deltaopt^\pm=-\Omega\cosh(2r)\pm\tfrac{1}{2}\sqrt{4\Omega^2\sinh^2(2r)-\kappa^2}}$, provided that the identification $r=r_{\mathrm{OM}}/2$ is made.

\subsection{Limits of The Cooling Enhancement}
This section considers several important effects that limit the cooling enhancement provided by the squeezing.
They will be presented in order of practical importance with respect to our experimental parameters.

\subsubsection{\label{ssec:model}Drive Field Impurity}
\begin{figure}
	\includegraphics[width =\columnwidth]{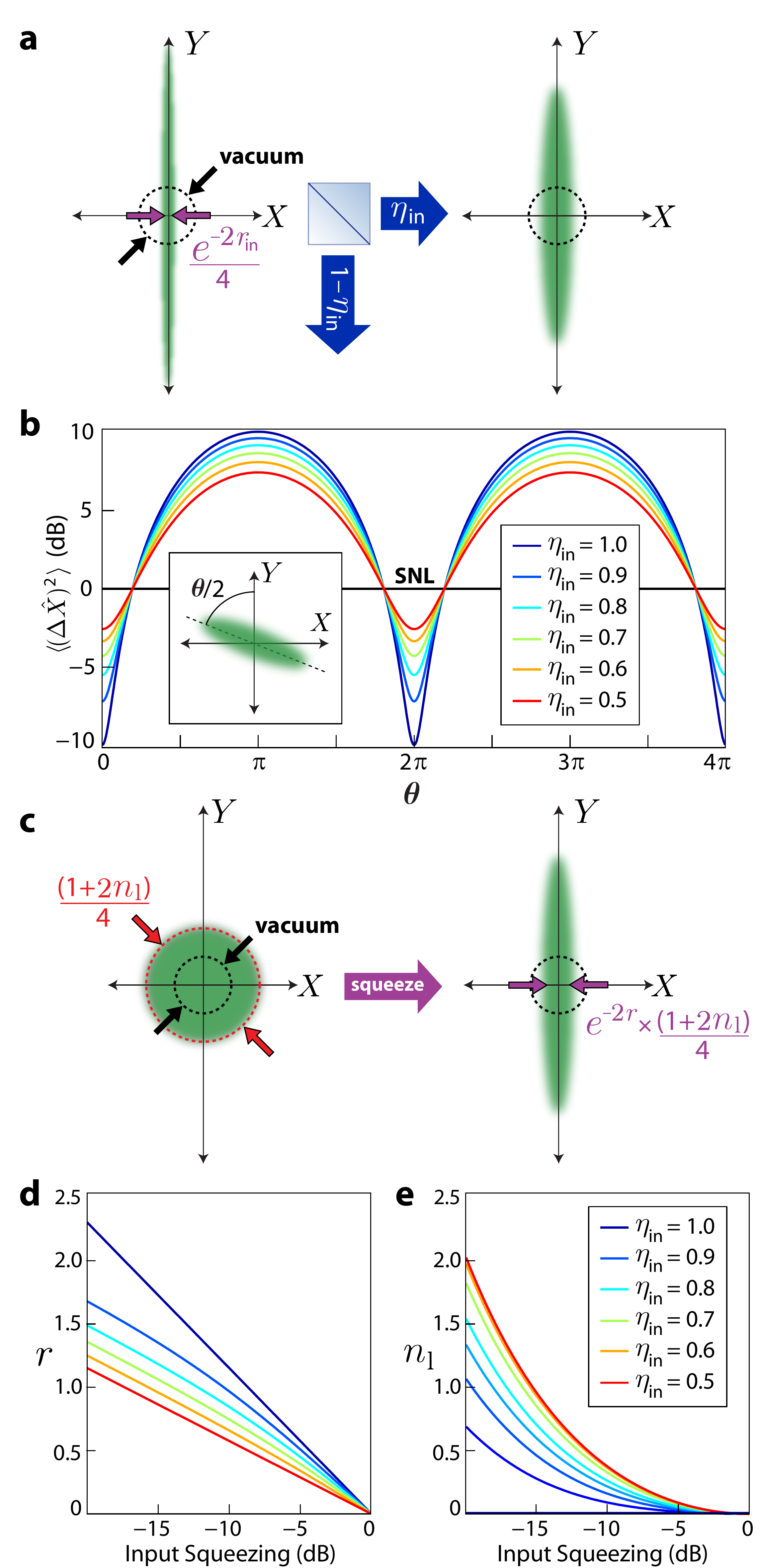}
	\caption{\label{fig:mcd}
		Modeling the impurity of the squeezing.
		\textbf{a}, A pure squeezed state (whose squeezing strength is parameterized by $\rin$) is subject to loss via a beamsplitter of transmittance $\etain$.
		Though the loss is to vacuum, the resulting squeezed state is impure and does not obey the minimum uncertainty relation: ${\braket{(\Delta\hat{X})^2(\Delta\hat{Y})^2}\neq\tfrac{1}{16}}$.
		\textbf{b}, Amplitude variance $\braket{(\Delta\hat{X})^2}$ of a pure ${-10~\mathrm{dB}}$ squeezed state subject to loss as the squeezing phase, $\theta$, is rotated (see inset).
		\textbf{c}, The Gaussian state on the right-hand side of (a) can be equivalently represented with $\nthd$ and $r$.
		The impurity is captured by the uncertainty product ${\braket{(\Delta\hat{X})^2(\Delta\hat{Y})^2}=(1+2\nthd)^2/16}$, which is maintained in the presence of an ideal entropy-preserving squeezing operation (parameterized by $r$).
		\textbf{d} and \textbf{e}, Computed values of $r$ and $\nthd$ given various values of $\rin$ and $\etain$ (see Eqs.~\ref{eq:r} and \ref{eq:nbar}).
	}
\end{figure}
In our experiments, the impurity of squeezed state driving the optomechanical system primarily limited the cooling enhancement.
From conventional homodyne tomography and optomechanical quantum nondemolition measurements of the light \cite{clark_observation_2016}, we conclude that the impurity can be largely explained by microwave loss in the critical path between the JPA and the optomechanical circuit (Fig.~\ref{fig:mcd}a).
Modeling the variance of the squeezed field's amplitude quadrature ${\hat{X}=\tfrac{1}{2}(\hat{a}^\dagger+\hat{a})}$ as a pure squeezed state subject to such loss yields
\begin{equation}
\label{eq:loss}
\braket{(\Delta\hat{X})^2}=\frac{1-\etain+\etain\times(\cosh 2\rin-\cos\theta\sinh 2\rin)}{4}.
\end{equation}
Here, $\theta$ denotes the squeezing phase, $\rin$ the squeezing parameter for the \textit{pure} squeezed state being injected into the critical path, and $\etain$ the critical path microwave transmittance (${\etain\approx57\pm2\%}$).
Figure~\ref{fig:mcd}b plots the variance as the squeezing phase, $\theta$, is rotated for several different values of $\etain$.

Modeling the squeezing in this way can be convenient since only a single parameter, $\rin$, is needed to specify a squeezed state subject to a given level of loss.
For other reasons, however, it can be more convenient to use the model described in the main text where the state is specified by an effective thermal photon occupancy, $\nthd$, and an ideal squeezing parameter, $r$.
This approach dramatically simplifies the expression for the full bath temperature $\nbathimpure$ in the weak coupling limit (Eq.~\ref{eq:nbathtot}).
% By extension, Eq.~\ref{eq:nbathtot} directly reveals how $\nbathimpure=\nthd$ when the pure squeezing expression $\nbath=0$ (Eq.~\ref{eq:nbath}).
Additionally, the loss model empirically shows poor agreement with our experimental data at higher inferred values of $\rin$.
In this regime, the squeezed state entering critical path can no longer be assumed pure.
Thus, at high squeezing strengths ($>$8~dB), two parameters are once again required to completely specify the quantum state of the drive field in our experiments.

After a thermal state of occupancy $\nthd$ is subject to an ideal squeezing operation, the amplitude quadrature variance obeys
\begin{equation}
\label{eq:param2}
\braket{(\Delta\hat{X})^2}=\frac{(1+2\nthd)\times(\cosh 2r-\cos\theta\sinh 2r)}{4}.
\end{equation}
Equation~\ref{eq:param2} confirms that the squeezing process is entropy-preserving since the uncertainty product of the maximally squeezed ($\theta=0$) and antisqueezed ($\theta=\pi$) quadratures is maintained for any value of $r$.
A Gaussian state that has been specified by $\rin$ and $\etain$ can be equivalently specified by $\nthd$ and $r$ according to
\begin{align}
\label{eq:r}r&=\ln\bigg[\bigg(\frac{1-\etain+\etain e^{2\rin}}{1-\etain+\etain e^{-2\rin}}\bigg)^{1/4}\bigg]\\
\label{eq:nbar}\nthd&=-\frac{1}{2}+\frac{1}{2}\sqrt{(1-\etain+\etain e^{-2\rin})(1-\etain+\etain e^{2\rin})}.
\end{align}
The behavior of $\nthd$ and $r$ for given values of $\rin$ and $\etain$ is plotted in Fig.~\ref{fig:mcd}d--e.
In our experiments, we found that the measured squeezing and the resulting cooling enhancement agrees well with the $\{r,\nthd\}$ pairs computed by Eqs.~\ref{eq:r} and \ref{eq:nbar} (assuming $\etain=57\%$), provided that $r<0.5$ (which corresponds to approximately $-8~\mathrm{dB}$ of injected squeezing).
As $r$ is increased further, the inferred values of $\nthd$ grow appreciably faster than the model predicts, suggesting the introduction of uncorrelated noise from the JPA.

\subsubsection{Strong Coupling}
It can be convenient to describe the optomechanical interaction in terms of cavity and mechanical susceptibilities, $\chi_{\mathrm{c}}$ and $\chi_\mathrm{m}$ (which are derived in the supplementary information of reference \cite{teufel_sideband_2011}).
From these susceptibilities, the following approximation is often made:
\begin{equation}
\label{eq:mech_approx}\frac{\chi_m}{1+g^2(\chi_c-\bar{\chi}_c)\chi_m}\approx\bigg(\Gamma_{\mathrm{tot}}/2-i(\omega-\Omega_{\mathrm{tot}})\bigg)^{-1},
\end{equation}
Under this approximation, the right-hand side of Eq.~\ref{eq:mech_approx} represents the ``dressed" mechanical susceptibility, which includes optical damping and springing effects.
The expressions for the mechanical bath occupancies in the main text (Eqs.~\ref{eq:nbath} and \ref{eq:nbathtot}) make use of this approximation.

When the applied drive power is increased to the point that the linearized coupling rate, $g$, exceeds the mechanical and cavity dissipation rates ($\Gamma$ and $\kappa$, respectively), the interaction enters the strong coupling regime.
In this regime, the approximation in Eq.~\ref{eq:mech_approx} is no longer valid.
The correction to the resolved sideband cooling limit imposed by strong coupling effects (\textit{e.g.} normal mode splitting) has been carefully studied for the case of coherent state drive fields \cite{dobrindt_parametric_2008, teufel_circuit_2011, verhagen_quantum-coherent_2012, wilson-rae_theory_2007}.
For a coherent state of the drive field, the final phonon occupancy, $\nf$, can be expanded in terms of the normalized parameters $\kappa/\Omega$ and $g/\Omega$ at zero temperature to yield \cite{dobrindt_parametric_2008, wilson-rae_theory_2007}
\begin{equation}
\label{eq:cooling_lim}
\nf\approx\bigg(\frac{\kappa}{4\Omega}\bigg)^2 + \frac{1}{2}\bigg(\frac{g}{\Omega}\bigg)^2.
\end{equation}
The first term in Eq.~\ref{eq:cooling_lim} corresponds to $\nmo$ (Eq.~\ref{eq:sblimit}) in the resolved sideband limit ($\kappa/\Omega\ll1$) when $\Delta=\Deltaopt$.
The $(g/\Omega)^2$ correction can be viewed as a strong coupling effect since it becomes comparable to $(\kappa/4\Omega)^2$ when $g\approx\kappa$.

\begin{figure}
	\includegraphics[width = \columnwidth]{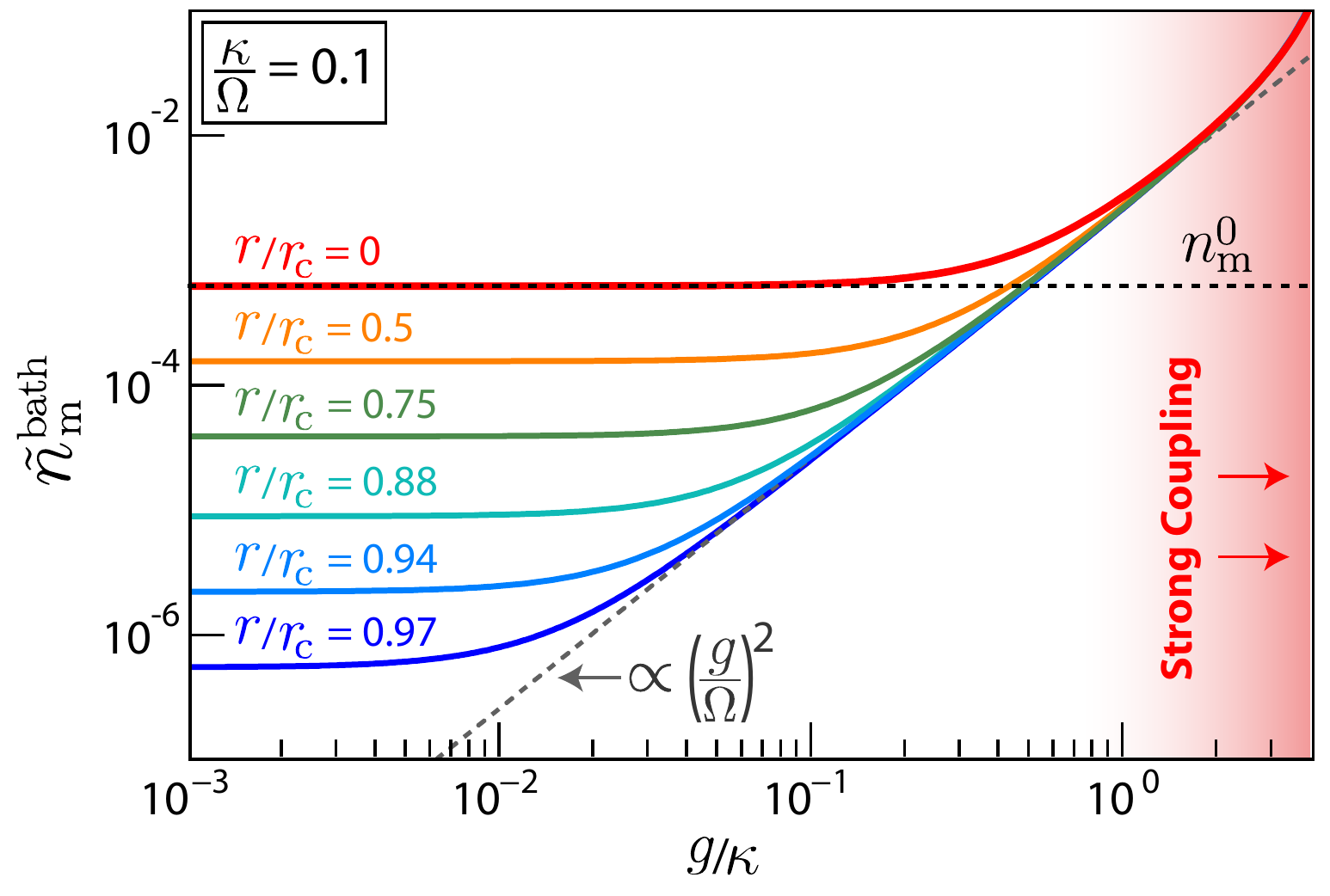}
	\caption{\label{fig:strong_coupling}
		The full numerically computed effective mechanical bath occupancy, $\nbathimpure$, is plotted as a function of the normalized coupling rate $g/\kappa$ for various pure squeezed states.
		By numerically computing the full bath occupancy, we account for the role of ``strong coupling" physics discussed elsewhere.
		Various squeezing parameters, $r$, have been considered (and are expressed in units of the critical squeezing parameter, $\rc$ in Eq.~\ref{eq:rc}).
		All curves assume drive detuning $\Delta=\Deltaopt$ (Eq.~\ref{eq:Deltaopt}) and squeezing phase $\theta=\thetaopt$ (Eq.~\ref{eq:thetaopt}).
		The optomechanical system is assumed to be deep in the resolved sideband cooling limit ($\kappa/\Omega = 0.1$).}
\end{figure}

Figure~\ref{fig:strong_coupling} generalizes this correction to the full mechanical bath occupancy $\nbathimpure$ resulting from using squeezed drive fields.
$\nbathimpure$ is numerically computed as a function of the normalized coupling strength, $g/\kappa$, assuming pure squeezed drive fields at the optimal squeezing phase ($\thetaopt$) and detuning ($\Deltaopt$).
A deeply sideband resolved optomechanical system ($\kappa/\Omega=0.1$) has been assumed.
The plot shows how $(g/\Omega)^2$ continues to bottleneck the cooling.
For the highest coupling rates in our experiments, however, this contribution to $\nbath$ falls two orders of magnitude below the heating effect arising from the impurity of the squeezing (see previous subsection).
Even when accounting for the ambient thermal environment in our experiments ($\nth\approx75$), the $(g/\Omega)^2$ contribution would in principle limit our experiments to $\nf\geq10^{-2}$ with pure squeezed states.
Therefore, it does not meaningfully limit our experimental results.

\begin{figure}
	\includegraphics[width =\columnwidth]{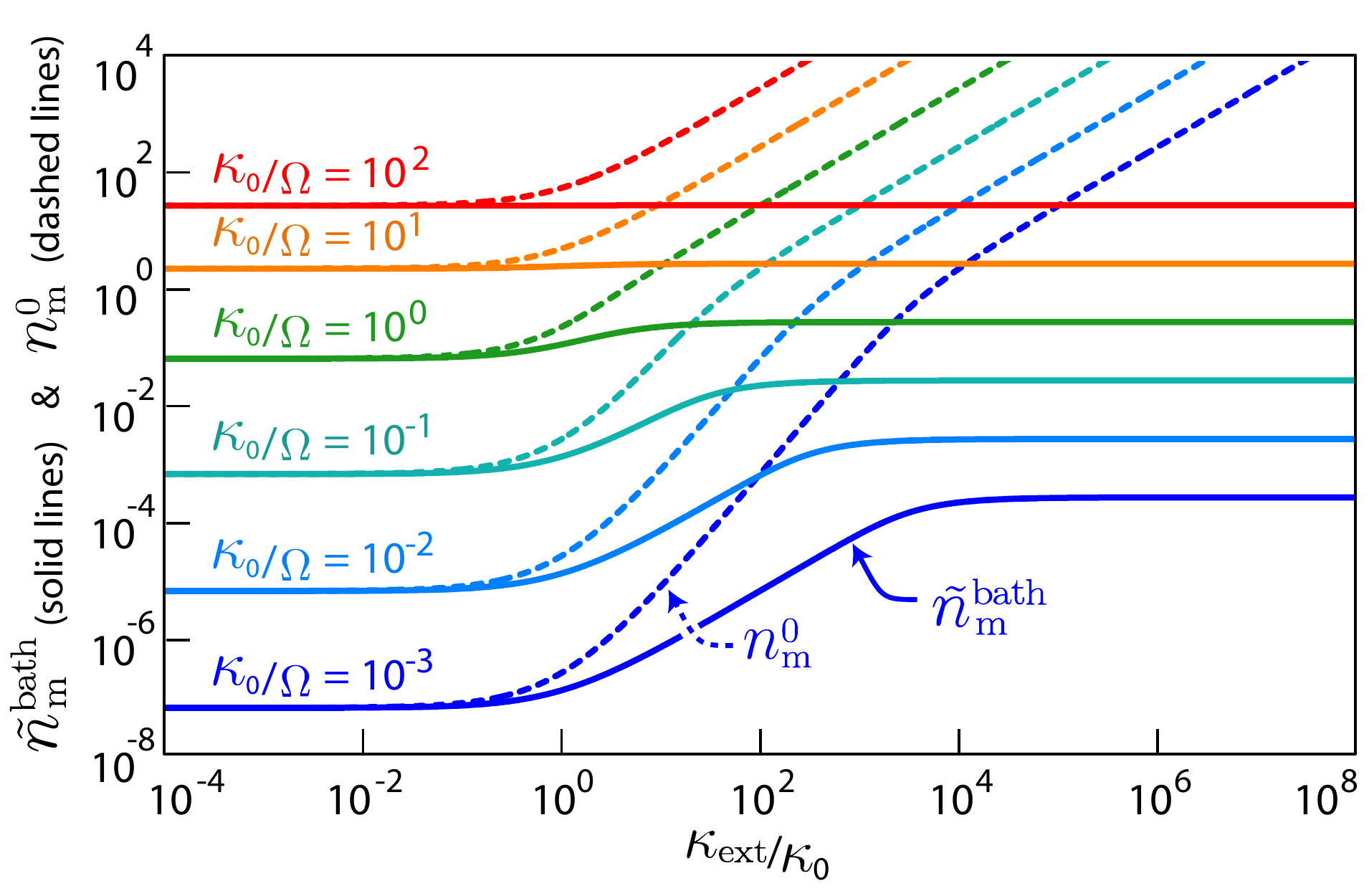}
	\caption{\label{fig:nbath_comparison}
		Comparison of the mechanical bath temperature with squeezed light ($\nbathimpure$, Eq.~\ref{eq:internal_limit}) and unsqueezed light ($\nmo$, Eq.~\ref{eq:sblimit}) in the presence of internal cavity loss.
		The bath occupancies are plotted as a function of the external coupling rate $\kex$ normalized to the cavity's internal loss rate, $\kappa_0$.
		Each color corresponds to a different value of $\kappa_0$.
		$\nbathimpure$ ($\nmo$) curves are denoted by solid (dashed) lines.
		All curves assume a drive detuning $\Deltaopt$ (Eq.~\ref{eq:Deltaopt}) from cavity resonance and the corresponding optimal squeezing phase $\thetaopt$ (Eq.~\ref{eq:thetaopt}).
		$\nbathimpure$ and $\nmo$ approach the same limit (Eq.~\ref{eq:limit_match}) as $\kex/\kappa_0\rightarrow0$, indicating that internal cavity loss limits the coldest temperatures that can be reached when sideband cooling with any Gaussian state.
		For $\kex\gg\kappa_0$, $\nmo$ diverges (Eq.~\ref{eq:nmo_lim}) whereas $\nbathimpure$ asymptotically approaches the ratio $\kappa_0/4\Omega$ (Eq.~\ref{eq:nbath_lim}).
	}
\end{figure}

\subsubsection{Cavity Loss}
In our experiments, the optomechanical circuit was strongly overcoupled to the feed line carrying the squeezed microwave field.
This ``external" coupling rate ${\kex\approx~27~\mathrm{MHz}}$ dominated the ``internal" cavity loss rate $\kappa_0<100~\mathrm{kHz}$ by at least two orders of magnitude.
Accordingly, microwave loss inside of the cavity did not play a significant role in limiting the achievable cooling enhancement.
Nevertheless, it is worth considering the effects of internal cavity loss since they become most important when operating in the desirable resolved sideband cooling limit.
Moreover, the effect of cavity loss is subtle; it cannot be correctly modeled as a loss channel outside of the cavity.

We will briefly describe how internal cavity loss limits sideband cooling for any Gaussian state of the drive field.
To simplify this discussion, we will neglect the strong coupling effects that were discussed in the previous subsection.
Additionally, to consider only fundamental limits, we will assume access to arbitrarily strong (pure) squeezed states at the input port of the cavity.
We will also assume that the cavity is held at low enough temperatures that the internal loss process can be treated as coupling to microwave vacuum.
Finally, we will focus the discussion on the behavior of the effective mechanical bath temperature in the presence of cavity loss, $\nbathimpure$ (as opposed to the final phonon occupancy, $\nf$ in Eq.~\ref{eq:nf}).

In describing how $\nbathimpure$ is affected by $\kappa_0$, it will be useful to define $\eta=\tfrac{\kex}{\kappa}=1-\tfrac{\kappa_0}{\kappa}$.
For any $0<\eta<1$,
\begin{equation}
\label{eq:internal_limit}
\nbathimpure(\kappa) = \eta\times\nbath(\kappa)+(1-\eta)\times\nmo(\kappa),
\end{equation}
where $\kappa = \kex + \kappa_0$.
In Eq.~\ref{eq:internal_limit}, $\nmo(\kappa)$ denotes the resolved sideband cooling limit for coherent states (Eq.~\ref{eq:sblimit}), and $\nbath(\kappa)$ represents the mechanical bath temperature in the presence of pure squeezing (Eq.~\ref{eq:nbath}).
It can be helpful to interpret $\nbathimpure(\kappa)$ as the result of coupling the mechanical mode to $\nbath(\kappa)$ and to $\nmo(\kappa)$ via a beamsplitter of transmittance, $\eta$.

By assuming access to pure and arbitrarily strong squeezing, it is always possible to choose a squeezing parameter $r$ such that $\nbath(\kappa)=0$.
Furthermore, by eliminating the first term in Eq.~\ref{eq:internal_limit}, it becomes clear that the drive detuning that minimizes $\nbathimpure(\kappa)$ is the same detuning, $\Deltaopt$, that minimizes $\nmo(\kappa)$.
We conclude that
\begin{equation}
\label{eq:nbath_crit}
\nbathimpure(\kappa)\geq\frac{\kappa_0}{\kex+\kappa_0}\times\bigg(\frac{1}{2}\sqrt{1+\bigg(\frac{\kex+\kappa_0}{2\Omega}\bigg)^2}-\frac{1}{2}\bigg),
\end{equation}
where the term in parenthesis gives the $\Delta=\Deltaopt$ expression for $\nmo(\kappa)$.
Since $\nbathimpure(\kappa)$ decreases as $\kex\rightarrow0$ (given some internal loss rate $\kappa_0$ being held constant), we conclude that
\begin{equation}
\label{eq:limit_match}
\nbathimpure(\kappa)>\nmo(\kappa_0).
\end{equation}
Equation~\ref{eq:limit_match} conveys that the cooling limit for \textit{any} Gaussian state is set by the cavity's internal loss rate.

Despite the limit imposed by $\kappa_0$, Fig.~\ref{fig:nbath_comparison} illustrates how pure squeezed light always provides a colder mechanical bath temperature than unsqueezed light given $\kex>0$.
The lower temperature bound given by Eq.~\ref{eq:limit_match} is asymptotically approached in the weakly undercoupled limit where ${\kex\ll\kappa_0}$.
On the other hand, in the strongly overcoupled limit (${\kex\gg\kappa_0}$), Eq.~\ref{eq:nbath_crit} yields a distinctly different behavior than that of $\nmo$:
\begin{align}
\label{eq:nbath_lim}\nbathimpure(\kappa)&\overset{\kex\gg\kappa_0}{\approx}\frac{\kappa_0}{4\Omega}\\
\label{eq:nmo_lim}\nmo(\kappa)&\overset{\kex\gg\kappa_0}{\approx}\frac{\kex}{4\Omega}.
\end{align}
The coherent state limit $\nmo(\kappa)$ diverges as $\kex\rightarrow\infty$, whereas $\nbathimpure(\kappa)$ asymptotes to a value given by the internal loss rate.
Thus, given a strongly overcoupled cavity, the cooling advantage conferred by a pure state of squeezed light increases as $\kex/\kappa_0$ grows.
Moreover, in the ``intrinsic" bad cavity limit where $\kappa_0/2\Omega\gg1$, $\nbathimpure(\kappa)$ becomes largely insensitive to changes in $\kex$ altogether.

\onecolumngrid
\section{Supplementary Tables}
\begin{figure*}[h!]
	\begin{center}
		\begin{tabular}{|l|c|c|}
			\hline
			\textbf{Parameter} & \textbf{Symbol} & \textbf{Value} \\
			\hline
			~~Mechanical resonance frequency & $\Omega$ & ~~$2\pi\times~10.1~\mathrm{MHz}$~~ \\
			\hline
			~~Total cavity linewidth & $\kappa$ & $2\pi\times~27~\mathrm{MHz}$ \\
			\hline
			~~Resolved sideband cooling limit (phonon occupancy) & $\nmo$ & $0.33$ \\
			\hline
			~~Critical squeezing parameter & $\rc$ & $0.55$\\
			\hline
			~~Cavity resonance frequency & $\omega_c$ & $2\pi\times~6.4~\mathrm{GHz}$ \\
			\hline
			~~Cavity internal loss rate & $\kappa_0$ & $<2\pi\times100~\mathrm{kHz}$ \\
			\hline
			~~Intrinsic mechanical linewidth & $\Gamma$ & $2\pi\times~15~\mathrm{Hz}$ \\
			\hline
			~~Vacuum optomechanical  coupling rate & $g_0$ & $2\pi\times$~260 Hz \\
			\hline
			~~Microwave critical path transmittance (between JPA and OM)~~& $\etain$ & $57\pm2\%$ \\
			\hline
			~~Input squeezing parameter (before critical path) & $\rin$ & - \\
			\hline
			~~\textit{Effective} thermal occupancy of the squeezing & $\nthd$ & - \\
			\hline
			~~\textit{Effective} entropy-preserving squeezing parameter & $r$ & - \\
			\hline
			~~Detection chain noise temperature~~ & $T_{\mathrm{N}}$ & $4.6~\mathrm{K}$\\
			\hline
			~~Effective heterodyne detection efficiency & $\eta_{\mathrm{det}}$ & 6.5\% \\
			\hline
			~~Base temperature of dilution cryostat & $T$ & 37 mK \\
			\hline
		\end{tabular}
		\caption{\label{table:constants}Summary of notation.  The corresponding values of any system constants are given.}
	\end{center}
\end{figure*}
\end{document}